# The first principle of neural circuit and the general Circuit-Probability theory

Hao Wang[1], Jiahui Wang[1], Xin Yuan Thow[2], Ng Kian Ann[2], Chengkuo Lee[1]

[1]Department of Electrical and Computer Engineering, National University of Singapore, Singapore 117583

[2]Singapore Institute for Neurotechnology (SINAPSE), National University of Singapore, Singapore 117456

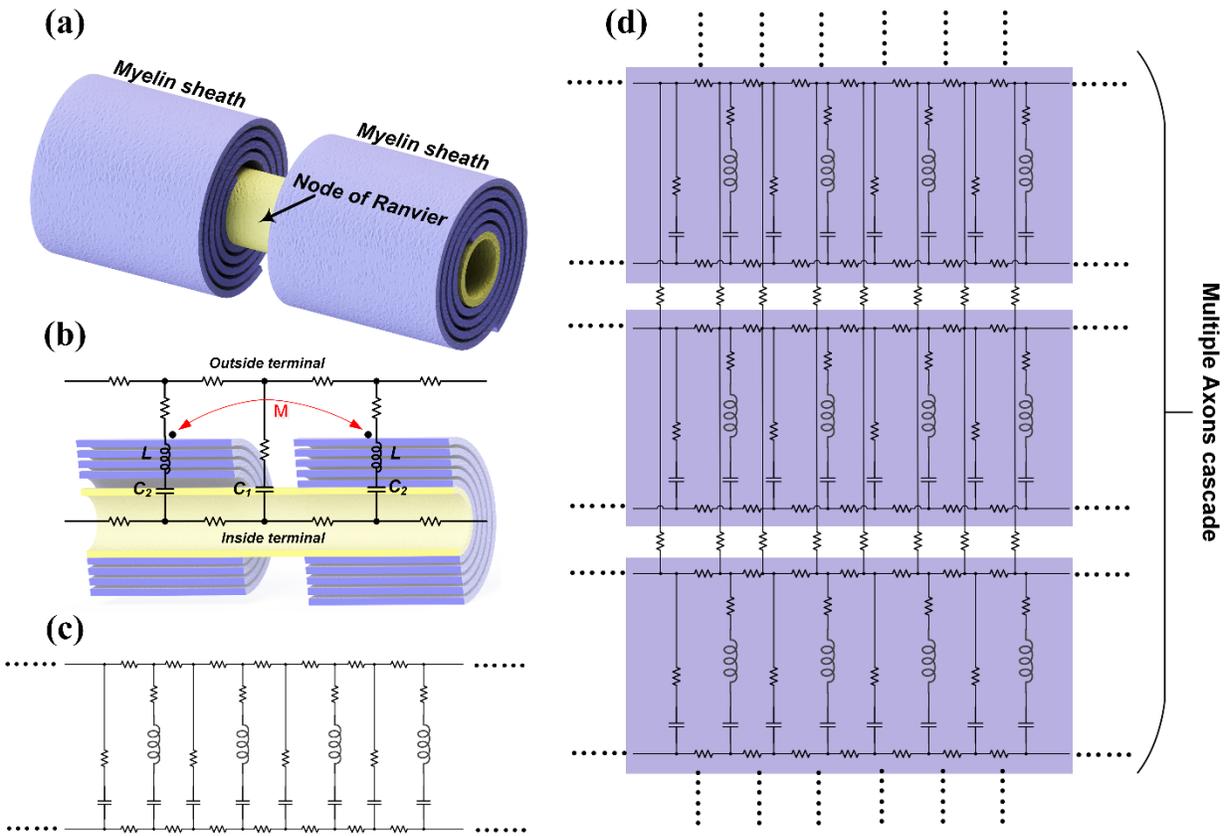

Figure 1. (a) 3D illustration of biological structure of a myelinated axon segment; (b) cross sectional view of the axon segment with the equivalent circuit model; (c) equivalent circuit cascade of a myelinated axon; (d) equivalent cascaded circuit network of multiple axons.

In the previous C-P theory [1], the myelin sheath is considered as the component to provide inductance in the lumped-parameter circuit. The parallel RLC configuration of the circuit is a conjecture which can fit the force mapping curves measured by experiments well. Then an inevitable question is why a complex nervous system can be simplified as a simple parallel RLC circuit.

To answer this question, a new distributed-parameter circuit model of a single axon, with the consideration of myelin sheath as an inductor, is proposed. This distributed-parameter circuit can be directly simplified as a lump-parameter parallel RLC circuit, which is proposed in the previous C-P theory. Moreover, this distributed-parameter circuit model also explains why the myelin sheath can enhance the propagation speed of a neural signal. Finally, this circuit model can explain many special phenomena in nerve stimulation by magnetic field and further provide guidance for the optimization of coil design.

The biological structure of a segment of myelinated axon is shown in Figure 1(a). The proposed distributed-parameter circuit model is as shown in Figure 1(b), which is equivalent to a segment of a single axon shown in Figure 1(a). This axon segment has two parts with different structures: part with myelin sheath and node of Ranvier without myelin sheath. For the Ranvier node, the cell membrane can be modeled as capacitors, $C_1$, connecting the inside and outside of the cell membrane. The intracellular and extracellular fluid can be modeled as resistors. For the part covered by myelin sheath, an extra inductors, $L$, is connected between the cell membrane, $C_2$, and extracellular fluid. Then the whole axon can be modeled as a circuit cascade as shown in Figure 1(c). Multiple axons can be modeled as a cascade network as shown in Figure 1(d). The outside terminals of each axon are connected with resistors.

Meanwhile, a key feather need to be added in this circuit, which is the mutual inductance between the adjacent myelin sheaths. In term of physics, this mutual inductance can be either positive or negative. In this circuit model, a definite conclusion will be drawn that this mutual inductance can only be positive (will be explained in the next section), meaning that the wrapping orientations of adjacent myelin sheaths have to be opposite. We noticed that this phenomenon was firstly observed in a study of schwann cell nucleus in 1989 [2]. After that no further research focused on studying this phenomenon since the myelin sheath is conventionally considered as merely a non-conductive layer. An intensive and comprehensive study of the wrapping orientation of myelin sheath can provide substantial evidence to support this theory in the future.

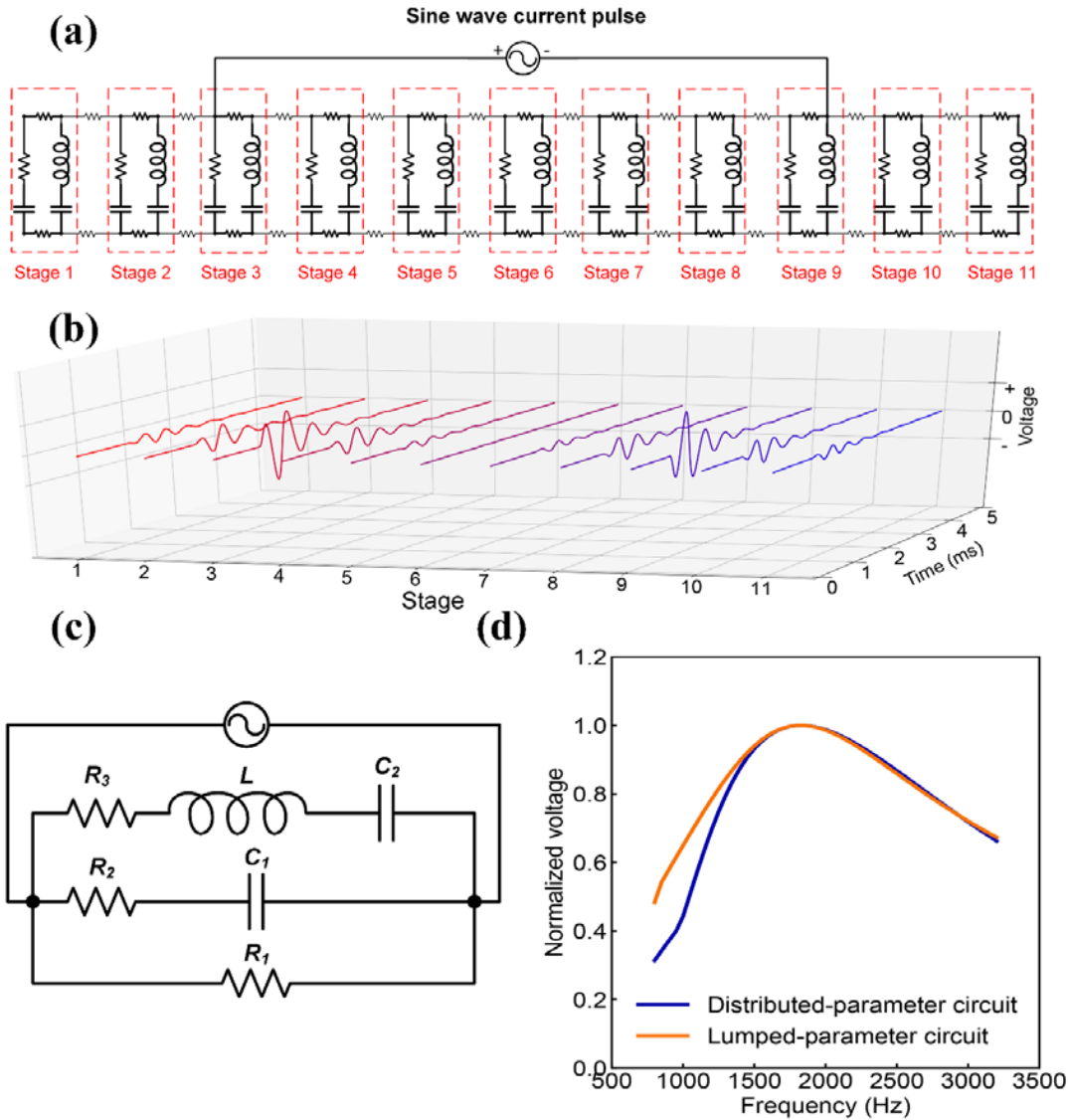

Figure 2 (a) A circuit cascade with 11 stages for modeling; (b) The voltage waveforms on $C_1$ of each stage, showing a gradual change of the amplitude the polarity reversing; (c) The simplified lumped-parameter circuit proposed in previous C-P theory; (d) The frequency response curve of the voltage amplitude on $C_1$ of the stage 4 in (a) and (c).

**How the distributed-parameter circuit compatible with the lumped-parameter circuit in C-P theory**

A single axon can be modeled as a cascade of basic units, indicated within the red dash box in Figure 2(a)). The configuration of this basic unit is a parallel RLC circuit with an extra capacitor ($C_2$) connected in series with the inductor. Here the physical meaning of $C_1$ is the capacitance of the Ranvier node and $C_2$ is the capacitance of the cell membrane covered by myelin sheath. Considering that the length of the segment covered by myelin sheath is much longer than the Ranvier node, here $C_2 \gg C_1$. Apparently, this cascade can be simplified as a circuit which has the same configuration as its basic unit, as shown in Figure 2(c).

This simplified circuit is exactly the revised lumped-parameter circuit proposed in C-P theory. In the initial version of the lumped-parameter circuit in C-P theory, the extra capacitor $C_2$ is not included. However, in our previous research, we have already shown that the force mapping curves can be better fitted with this extra $C_2$. Now based on the distributed-parameter circuit, which is derived from the biological structure of the axon, we know this $C_2$ really exists and its physical meaning is the cell membrane covered by myelin sheath.

This circuit cascade can be directly used for the study of the nerve stimulation. An equivalent circuit cascade with 11 stages is shown in Figure 2(a). A current source is connected to the stage 3 and 9 (Figure 2(a)), which is the similar situation as an electrode pair implanted on a nerve branch. When a sinewave current pulse is applied, the voltage waveforms on the capacitors, $C_1$, of each stage are shown in Figure 2(b). As can be seen, the voltage waveforms will gradually change from a positive-first phase to a negative-first phase. It means the segment close to the positive electrode actually has a different voltage waveform with the segment close to the negative electrode. As a simple prediction, with a same stimulating current waveform, the stimulation to the upstream and downstream will be different. With correct circuit parameters, this difference shall be well explained by the polarity reversing of the voltage waveform. ***The detailed modeling parameters can be found in Table 1(?).***

Then a typical case demonstration in Figure 2(d) shows how the lumped-parameter circuit (Figure 2(c)) generates the same electric characteristics as the distributed-parameter circuit (Figure 2(a)), validating the circuit simplification proposed in the previous C-P theory. With the same configuration as shown in Figure 2(a), by varying the frequency of the sinewave current pulse while keeping the amplitude the same, the frequency response of the voltage amplitude on $C_1$ of stage 4 is shown as the blue curve in Figure 2(d), which is a typical frequency response curve of a parallel RLC circuit. Then by tuning the circuit parameters in Figure 2(c), a similar frequency response of the voltage amplitude on $C_1$ in the lumped-parameter circuit (Figure 2(c)) is shown as the orange curve in Figure 2(d), which is almost the same as the blue curve.

In summary, if we only care about the voltage waveforms on the Ranvier node, which determines the result in electrical nerve stimulation, a lumped-parameter circuit in Figure 2(c) can simplify the analysis. However, an in-depth study of the distributed-parameter circuit can help understand more about the nervous system, which will be explained in the following sections.

**How the myelin sheath enhance the propagation speed of neural signals**

In previous neural model, the myelin sheath is modeled as a pure resistor to increase the propagation speed of neural signals by decreasing capacitance and increasing electrical resistance across the cell membrane.

However, in our new circuit model by considering the myelin sheath as an inductor, this propagation speed enhancement will have a new explanation.

Generally speaking, this propagation speed enhancement is in induced by two mechanisms: mutual inductance between two adjacent myelin sheaths and frequency modulated signal decay. To demonstration these two mechanism, a circuit which is similar to one in Figure 2(a) is proposed in Figure 3(a), with a different configuration of the current source. The current source is connected in parallel with the $C_1$ of stage 6. A sine wave current pulse in applied to mimic the generation of action potential.

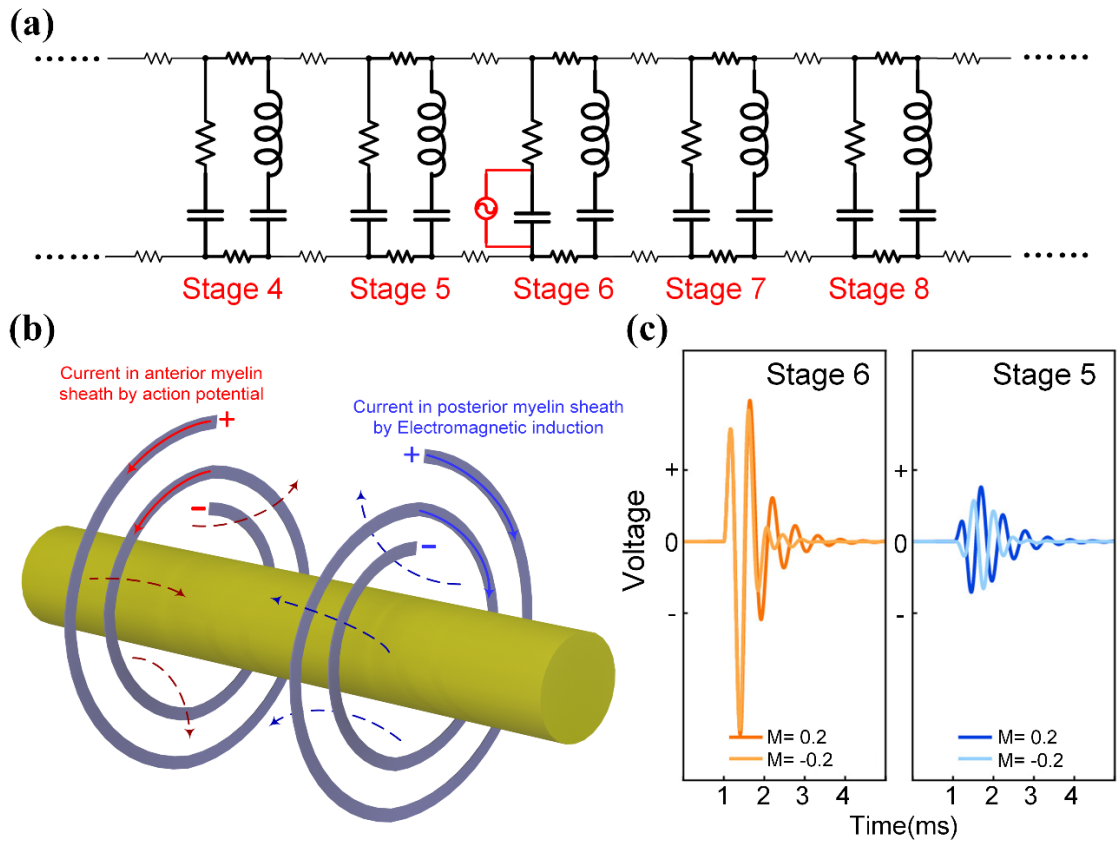

Figure 3 (a) A circuit cascade with 11 stages for modeling, the current source for applying a sine wave current pulse is connected on $C_1$ of stage 6 to mimic the activation of an action potential; (b) Illustrative plotting of the mutual inductance between adjacent myelin sheaths; (c) Comparison of the voltage waveforms on $C_1$ of stage 5 and 6 with positive and negative mutual inductance.

**Mutual inductance**

Since the node of Ranvier is very short, the mutual inductance between the two adjacent myelin sheaths will not be negligible. An illustrative plotting for this mutual inductance is shown in Figure 3(b). When an action potential is activated on the node of Ranvier before the anterior myelin sheath, this action potential is also applied onto this myelin sheath. Assuming at an instant, the potential cross the myelin sheath is positive outside and negative inside, then a current will flow along the wrapping orientation from outside to inside, generating a magnetic field. This magnetic field will generate an inductive current on the posterior myelin sheath. Due to the opposite wrapping orientations of adjacent myelin sheaths, the voltage induced by the inductive current on the posterior myelin sheath will share the same polarity, which is also positive outside and negative inside. This voltage will be further coupled onto the node of Ranvier between these two myelin sheaths. Surely, even without this mutual inductance, the action potential on the anterior node of Ranvier still can generate some voltage onto the posterior node of Ranvier. However, this mutual inductance can increase the voltage amplitude, reduce the signal decay and finally enhance the propagation speed of neural signal.

The effect of the mutual inductance upon the propagation of the action potential is shown in Figure 3(c). A sine wave current pulse is applied on $C_1$ of stage 6 to mimic the generation of an action potential. The voltage waveforms on $C_1$ of stage 6 and stage 5 are compared when the mutual inductance is positive (opposite wrapping orientations) and negative (same wrapping orientations). This mutual inductance only has a minor effect on the voltage on stage 6, but affects the voltage on stage 5 a lot. The voltage of positive mutual inductance is higher than that of negative mutual inductance. Meanwhile, the positive mutual inductance can guarantee a same voltage polarity between the anterior and posterior stage, which can make the activation of the action potential on the posterior stage earlier. But the negative mutual inductance may induce a polarity reversing. As shown in Figure 3(c), the voltage on stage 6 has a positive-first polarity while the voltage for negative mutual inductance on stage 5 has a negative-first polarity, which will delay the activation of action potential.

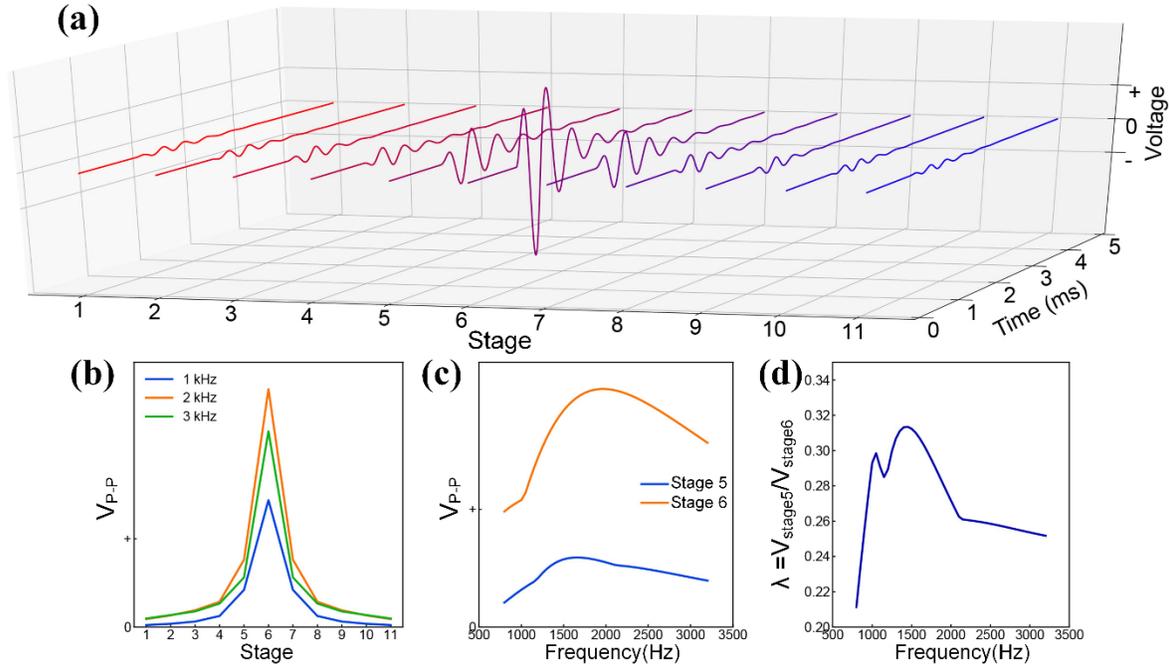

Figure 4 (a) Voltage waveforms on $C_1$ of each stage when an action potential is activated on stage 6, the circuit for modeling is shown in Figure 3(a); (b) The amplitude of the voltage waveform on each stage with different frequency of the applied sine wave current pulse applied on stage 6; (c) The voltage-frequency curve on $C_1$ of stage 6 and stage 5; (d) the curve of decay constant $\lambda = V_{stage5}/V_{stage6}$ with respect to frequency

**Frequency modulated signal decay**

Another effect induced by myelin sheaths upon the neural signal propagation is the frequency modulated signal decay. Considering the same situation as shown in Figure 3(a), an action potential is activated on the stage 6, the voltage waveforms on $C_1$ of each stage are shown in Figure 4(a). The voltage amplitude reaches maximum on stage 6 and gradually decay for the stages farther from stage 6. In this circuit, the decay constant $\lambda$ is defined as the voltage ratio between the anterior and posterior stages:

$$\lambda = V_{stage5}/V_{stage6}$$

This decay constant $\lambda$ is determined by the frequency of the input signal. A detailed circuit analysis can be found in **Supplement 1**. A simple conclusion drawn from the circuit analysis is that the decay constant $\lambda$ reaches maximum at the resonance frequency of each stage, meaning a minimum signal decay. This conclusion is also confirmed by modeling as shown from Figure 4(b)-(d). The voltage amplitudes of each stage with different input frequency is shown in Figure 4(b). For the curve of 2 kHz, which is the resonance frequency set for each stage, the response voltage has the maximum voltage and minimum decay. In Figure 4(c), the voltage-frequency curves of stage 5 and stage 6 are shown. These two curves all exhibit a parallel RLC circuit characteristic, which is similar to the result in Figure 2(d). Then calculate the decay constant $\lambda$ in Figure 4(d). As can been seen, $\lambda$ reaches maximum not at the resonance frequency set for modeling frequency, which is 2 kHz. This is because the resonance frequency set for modeling is calculated by the equation:

$$f = \frac{1}{2\pi\sqrt{LC_1}}$$

However, the actual resonance frequency is also affected by factors such as $C_2$, other resistors and mutual inductance. So the frequency with maximum $\lambda$ has a certain shift with 2 kHz. In this case, the actual resonance frequency is 1500 Hz. It means the action potential can propagate with a minimum decay when its major frequency is at 1500 Hz. Or in other words, the growth of the myelin will fit the major frequency of the action potential to realize a minimum signal decay.

In summary, the myelin sheath can enhance the propagation of neural signal by two mechanisms. The opposite wrapping orientations of adjacent myelin sheaths can induce a positive mutual inductance to increase the coupling between the anterior and posterior nodes of Ranvier, enhancing the propagation speed of action potential. The myelin will grow to a certain length and thickness to make the resonance frequency of the axon the same as the major frequency of the action potential, minimizing the signal decay along the axon to enhance the propagation speed of action potential.

**Nerve stimulation by magnetic field**

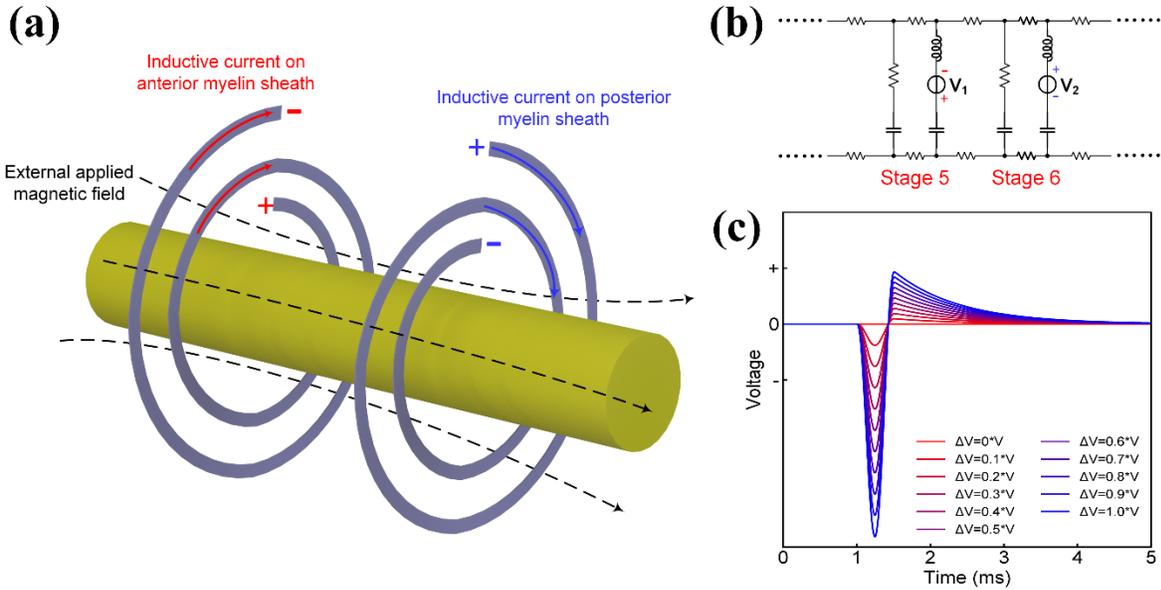

Figure 5 (a) Illustrative drawing of the inductive current on adjacent myelin sheaths by external applied magnetic field; (b) Equivalent circuit with voltage sources connected to adjacent stages; (c) The voltage on $C_1$ of stage 6, representing the voltage on the node of Ranvier between two myelin sheath, by increasing the voltage difference between two stages.

Since the myelin sheath is an inductor in our theory, an external applied magnetic field can also generate inductive current within the myelin sheath and further generate voltage on the node of Ranvier to activate action potential. An illustrative drawing of this electromagnetic induction is shown in Figure 5(a). When a magnetic field is applied along the longitudinal direction of the axon, the inductive current on the anterior and posterior myelin sheaths will flow in opposite directions because of their opposite wrapping orientations. So the inductive potential on these two myelin sheaths will have opposite polarity. Since the outside terminals of these two myelin sheaths are connected, these two inductive potential will cancel with each other and finally only their potential difference can be coupled onto the node of Ranvier in between. This phenomenon can be modeled by the circuit as shown in Figure 5(b). The inductive potential can be modeled as a voltage source connected in series with the inductor. Here only stage 5 and 6 are used for a simple demonstration. The voltage source of these two stages will apply sine wave current pulse with opposite polarities. The amplitudes of these two voltage sources are $V_1$ and $V_2$ for stage 5 and 6, respectively.

The difference between $V_1$ and $V_2$ is $\Delta V$:

$\Delta V = V_1 - V_1$;

The voltage on $C_1$ of stage 6, which is the node of Ranvier between two myelin sheaths of stage 5 and 6, by increasing $\Delta V$ is shown in Figure 5(c). As can be seen, the voltage on $C_1$ is proportional to $\Delta V$. When there is no potential difference, the voltage on $C_1$ is also zero.

Since the inductive potential on the myelin sheath is proportional to the amplitude of the magnetic field component in the longitudinal direction of the axon, the voltage difference between two adjacent myelin sheaths is proportional to the amplitude difference of this magnetic field, which is the field gradient along the axon. It means, it is the gradient, not the amplitude, of the magnetic field actually taking effect for magnetic nerve stimulation. In normal condition when a coil is positioned near a straight axon, the gradient of the magnetic field along the axon is very low. It explains why the current required for magnetic field stimulation is so high.

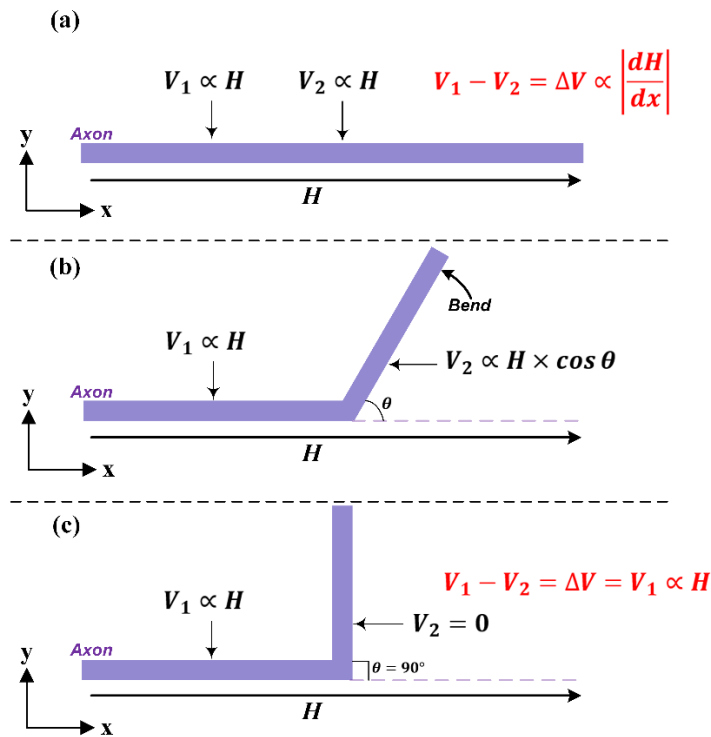

Figure 6 A detailed analysis of magnetic nerve stimulation with different axon bending angles. (a) The axon is straight; (b) The axon is bent with angle $\theta$; (c) The bending angle $\theta = 90°$.

A more detailed analysis is shown in Figure 6. Here only consider the situation when the magnetic field is along the axon.

When the axon is straight (Figure 6(a)), the inductive potential of the two near location is denoted as $V_1$ and $V_2$. Since the magnetic field is along the axon, these two potential are all proportional to the amplitude of the magnetic field. Their difference is proportional to the gradient of the magnetic field:

$$\Delta V \propto \left|\frac{dH}{dx}\right|$$

The direction along the axon is set as $x$ axis.

When part of the axon is bent with an angle $\theta$ as shown in Figure 6(b), the inductive potential $V_2$ will change as:

$$V_2 \propto H \times \cos\theta$$

Generally, $V_2$ will decrease since $\cos\theta \leq 1$, and $\Delta V$, the difference between $V_1$ and $V_2$, will increase, lowering the threshold current required for magnetic nerve stimulation.

When the bending angle $\theta = 90°$ as shown in Figure 6(c), $V_2 = 0$. Then the voltage difference $\Delta V = V_1$, which is the theoretical maximum value can be achieved. In this situation, the current required for magnetic stimulation will be minimum.

The analysis results for these three situations in Figure 6 have already been confirmed by previous study [3]. Now all these phenomena can be theoretically explained in our theory.

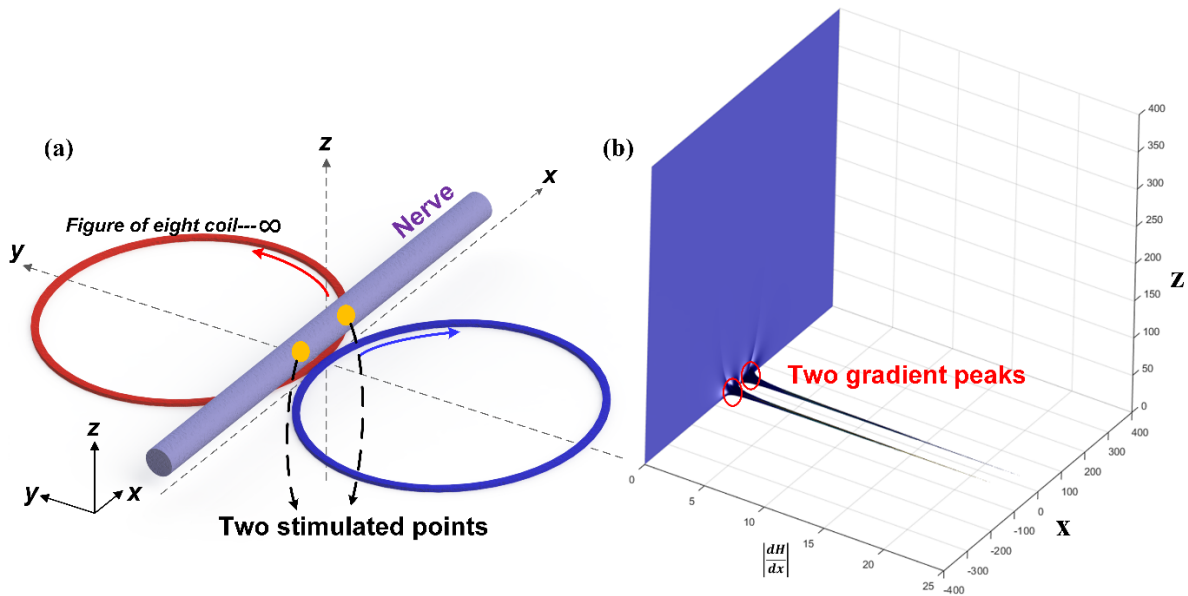

Figure 7 Modeling of magnetic field nerve stimulation by a figure of eight coil. (a) 3D illustrative drawing showing that a straight nerve along x axis is placed above a figure of eight coil, the current directions in two circles are anticlockwise and clockwise, respectively; (b) The modeling result of $\left|\frac{dH}{dx}\right|$, the gradient of the magnetic field component along x axis, in x-z plane, showing two peaks corresponding to the two stimulation points observed in previous study [3].

Here a simple case demonstration in Figure 7 shows how our theory predicts an exact experiment observation in previous study [3]. It was observed that there will be two stimulation points when a straight nerve is placed above the coil with figure of eight shape. Figure 7(a) shows the experiment configuration. A straight nerve along the x axis is placed above the coil and the current directions in two circles are anticlockwise and clockwise, respectively. Two yellow points refer to the two stimulation points. The magnetic field gradient along x axis, $\left|\frac{dH}{dx}\right|$, in x-z plane when y deviates a bit from zero point, is shown in Figure 7(b). As can be seen, there will be two peaks, corresponding to the two stimulation points observed in previous study.

It needs to be emphasized that the magnetic field along x axis is always zero when y=0, because of the perfect symmetric coil shape used in our modeling. The magnetic field along x axis generated by two circles will completed balance with each other. However, this situation will never happen in real experiment since the coils shape will never be perfectly symmetric and the nerve has a certain volume.

Based on the above explanation, it is clearly shown that the coil structure of myelin sheath can generate an inductive effect to realize an energy conversion between electric field and magnetic field. In perspective of circuit, myelin sheath should be considered as a real inductor, which can generate magnetic field from current and sense magnetic field by generating inductive current. However, based on the equation of the resonance frequency:

$$f = \frac{1}{2\pi\sqrt{LC}}$$

the inductance will be a huge value, can be as high as several Henry (H), when we consider the cell membrane capacitance as a reasonable value, which is of nF level, and the measured resonance frequency $f$ is at kHz level. This huge inductor has been measured in the original paper of H-H model and Cole's early stage results []. Apparently, the myelin sheath, which is a microscale object, cannot provide such as a huge inductance. Thus, there should be another mechanism to provide this huge inductance.

Actually, this huge inductor is not a real one, but an equivalent one. It is induced by the piezoelectric effect of the cell membrane. This piezoelectric effect is illustrated in Figure 8.

The cell membrane is of a lipid bilayer structure as shown in Figure 8(a). The lipid molecules are polar molecules, with positive tails toward the center and negative tails toward the extra- and intracellular fluid (shown as the + and - signs). Then when an external electric field is applied onto this lipid bilayer membrane, the two layers will have an opposite deformation as shown in Figure 8(c). When the electric field is top positive and bottom negative (shown in (c-i)), the top layer will be stretched and bottom layer will be compressed. When the electric field is top negative and bottom positive (shown in (c-ii)), the top layer will be compressed and bottom layer will be stretched. Here in the Figure 8(c), only axial deformation is depicted. However, the actual deformation will happen along all directions, resulting in a change of the surface stress.

This phenomenon is called piezoelectric effect in MEMS (Microelectromechanical systems). The equivalent circuit of this piezoelectric layer is shown in Figure 8(b). In the right branch, $C_P$ is the capacitance of the cell membrane, determined by its area, thickness and dielectric constant. Then in the left branch, the three parameters, $L$, $C$ and $R$, are just modeling parameters to fit the experiment result. They don't have exact physical meanings. Here The value of $C$ must be much smaller than $C_P$.

As can be seen, there will be an inductor in the equivalent circuit. The value of the inductance is determined by the capacitance of the cell membrane $C_P$ and the resonance frequency $f$. The resonance frequency is determined by the mechanical structure of the cell membrane. Considering the cell membrane is very thin and soft, it can be modeled as a very long, very thin and very soft beam, which has a very low mechanical resonance frequency. In our previous experiment on myelinated nerves, this resonance frequency is between 1k to 2k Hz. In Cole's early experiment on giant squid axon, which is a non-myelinated, this resonance frequency is around 300 Hz. Now it is easy to comprehend why the resonance frequency is so low. With this very low resonance frequency, a huge inductor in the circuit is inevitable, since it is just an equivalent inductor to fit the experiment result. This equivalent inductor will not induce any magnetic field, but introduce a mechanical surface stress.

Then as an easy prediction, this mechanical stress can also be measured as a mechanical wave, which is accompany with the propagation of the action potential. Since this mechanical wave is directly induced by the electric signal, the mechanical wave will also have the same propagation speed as the electric signal of the action potential.

This mechanical wave has been measured and studied for more than 10 years []. A theory, called soliton theory was developed based on the observation of this mechanical wave. Based on the calculation of soliton theory, the propagation speed of the mechanical wave should the same as the action potential. However, after understanding the physical mechanism of this mechanical wave, we can immediately know this even without any calculation.

Moreover, this piezoelectric effect of the cell membrane also explains the mechanism of acoustic nerve stimulation. Previous, there are several different possible mechanisms are proposed to explain why ultrasound can stimulate the nerve []. Now since the cell membrane is a piezoelectric film, the surface stress can be generated by applying electric field, and on the other hand, the surface stress can also generate electric field. The external applied acoustic wave can introduce this surface stress and then generate a voltage on the cell membrane to open the ion channels. This mechanism is totally the same as electric stimulation. The only difference is that, this voltage on the cell membrane is not induced by electrical coupling but by piezoelectric effect.

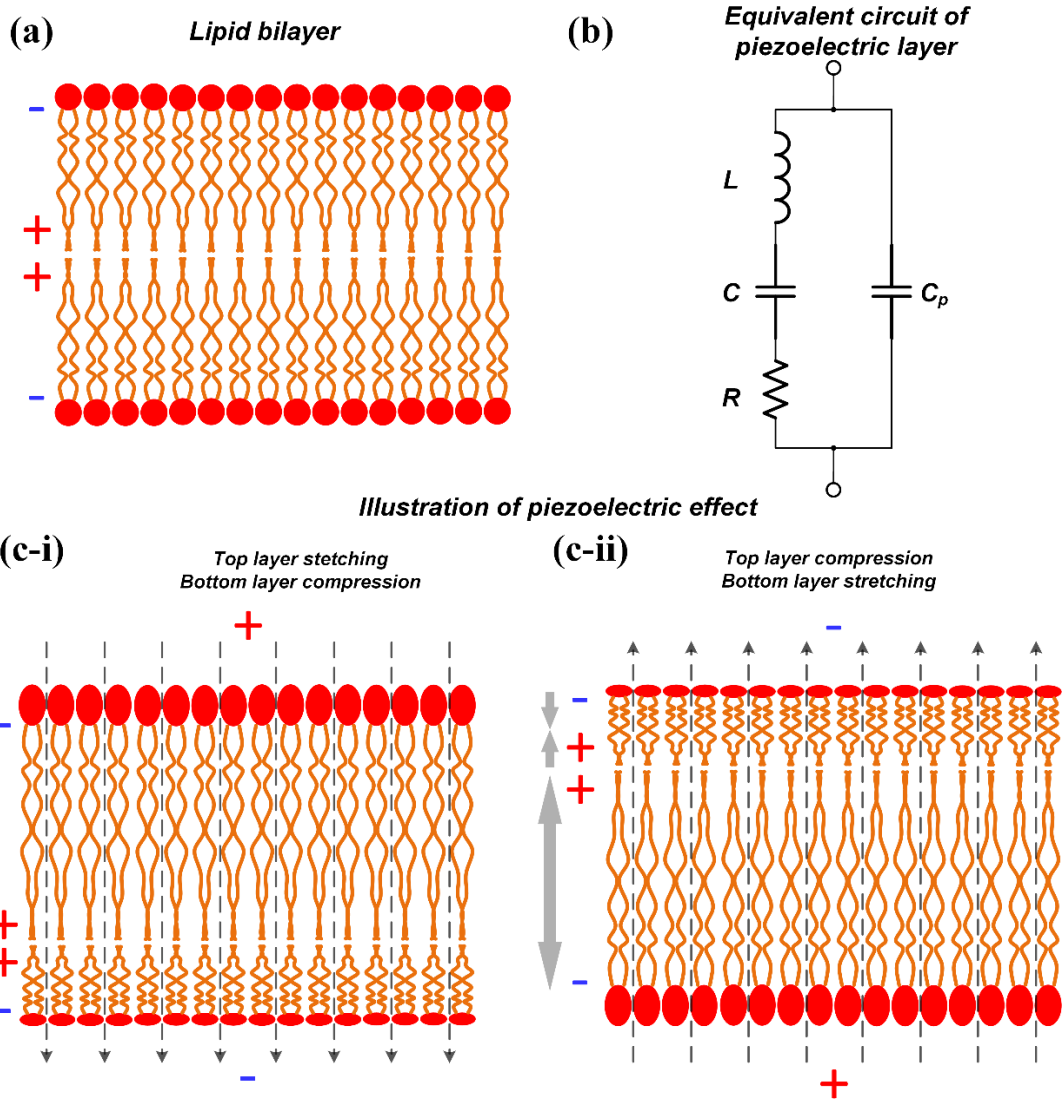

Figure 8 Illustration of the piezoelectric effect of the lipid bilayer structure of the cell membrane.

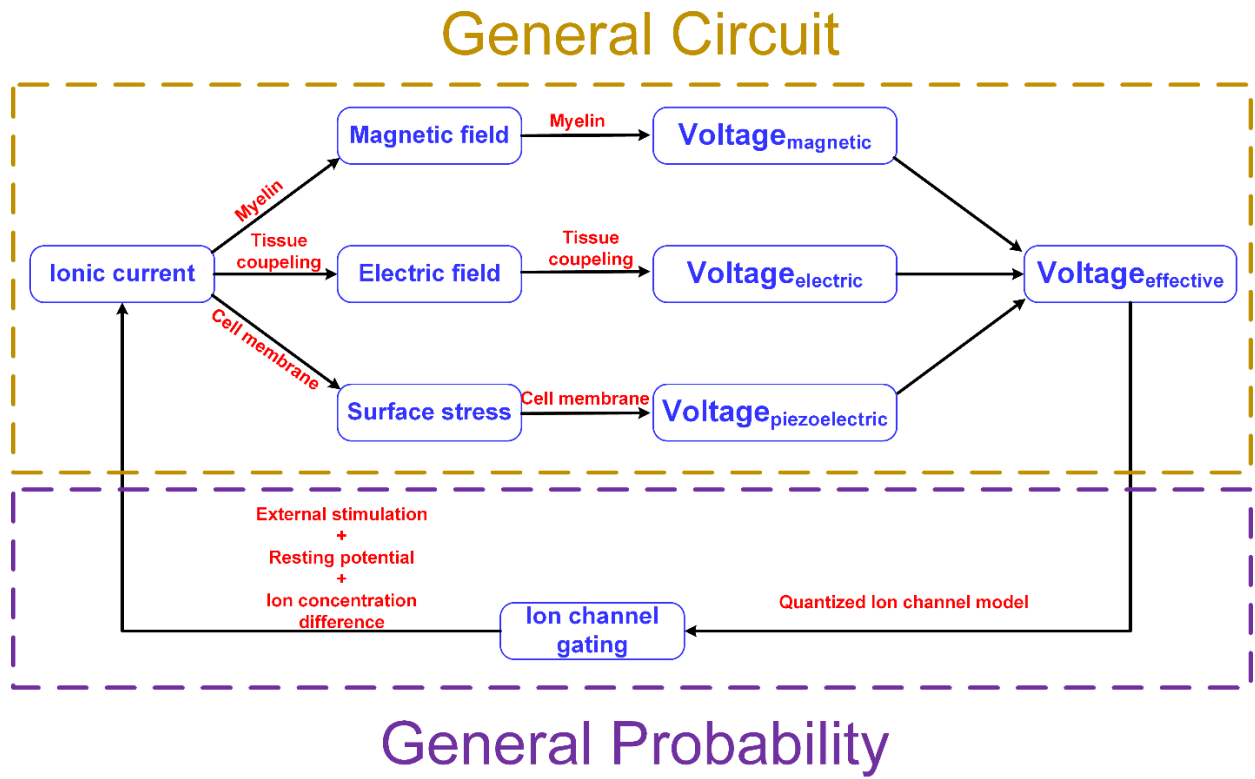

Figure 9 General C-P theory.

Previously, there is an argument about whether the neural signal is an electric signal or a mechanical signal. In H-H model, the activation of the action potential is explained in an electrical perspective. However, soliton theory provide an alternative illustration based on mechanical perspective. Here we provide a new platform, called general C-P theory, in the perspective of all first physical principles, to end this argument.

The mistake for both H-H model and soliton theory is that they mix the propagation and activation of the action potential together. This is because both of these two models are not developed based on first physical principles. However, in our general C-P theory, together with the quantized ion channel model [], every physical procedure involved in the activation and propagation of the action potential is now clear. Apparently, the physical mechanisms dominating the action potential propagation (explained by the general circuit) and activation (explained by the general probability) are totally different. The general circuit can be well described by classical physics while the general probability is described by quantum physics. It is impossible to combine these two parts together in one analytical model, which is the target pursued by both H-H model and soliton theory.

The procedure of neural signal can be divided into two general blocks, which are general circuit and general probability.

In general circuit, the ionic current induced by the gating of ion channels provides energy. This energy is converted into three forms: magnetic field by myelin, electric field by the tissue coupling and mechanical stress by the piezoelectric cell membrane. Then again, the energy of these three forms can be converted to electric voltage on the next node of Ranvier. By modeling all these three mechanisms by circuit, these three voltage can be summed up by a certain coupling, resulting in an effect voltage, $\boldsymbol{Voltage_{effective}}$. Here we called the circuit as general circuit because not everything involved is real electrical circuit. The magnetic coupling can be modelled by introducing mutual inductance, as shown in Figure 1. The piezoelectric coupling can be modelled as an equivalent circuit as shown in Figure 8(b). But the complete version of this general circuit is still unknown. This is because the myelin also provides a large cell membrane area, which means myelin has inductance and piezoelectric effect simultaneously. The equivalent circuit of the myelin sheath cannot be acquired by reasoning but only by experiment data fitting. A general data fitting suggests the circuit shown in Figure 2(c) is a good approximation. But the inductance involved should a combination of two parts: real electrical inductance by coil structure of myelin sheath and equivalent inductance by piezoelectric effect of cell membrane.

In general probability, the gating of the ion channel can be induced by this $\boldsymbol{Voltage_{effective}}$. This part is explained in our previous study []. Then based on the gating pattern, with the boundary condition of the external stimulation (electrical, magnetic or acoustic), resting potential and the ion concentration different of $\boldsymbol{Na^+}$ and $\boldsymbol{K^+}$, the ionic current can be calculated.

This general C-P theory provides a basic platform for understanding the activation and propagation of action potential. Meanwhile, it explains all physical mechanisms for nerve stimulation. The electrical stimulation provides the voltage by electrical coupling. The magnetic stimulation provides the voltage by electromagnetic induction of myelin sheath. The acoustic stimulation provides the voltage by piezoelectric effect of cell membrane. The optogenetics can directly open the ion channel by photoelectric effect.

**Reference**

[1] Hao Wang, Jiahui Wang, Xin Yuan Thow, Sanghoon Lee, Wendy Yen Xian Peh, Kian Ann Ng, Tianyiyi He, Nitish V. Thakor, Chengkuo Lee, Unveiling Stimulation Secrets of Electrical Excitation of Neural Tissue Using a Circuit Probability Theory, arXiv:1804.11310

[2] Bunge, R.P., Bunge, M.B. and Bates, M., 1989. Movements of the Schwann cell nucleus implicate progression of the inner (axon-related) Schwann cell process during myelination. The Journal of Cell Biology, 109(1), pp.273-284.

[3] Maccabee, P.J., Amassian, V.E., Eberle, L.P. and Cracco, R.Q., 1993. Magnetic coil stimulation of straight and bent amphibian and mammalian peripheral nerve in vitro: locus of excitation. The Journal of Physiology, 460(1), pp.201-219.

**Circuit analysis**

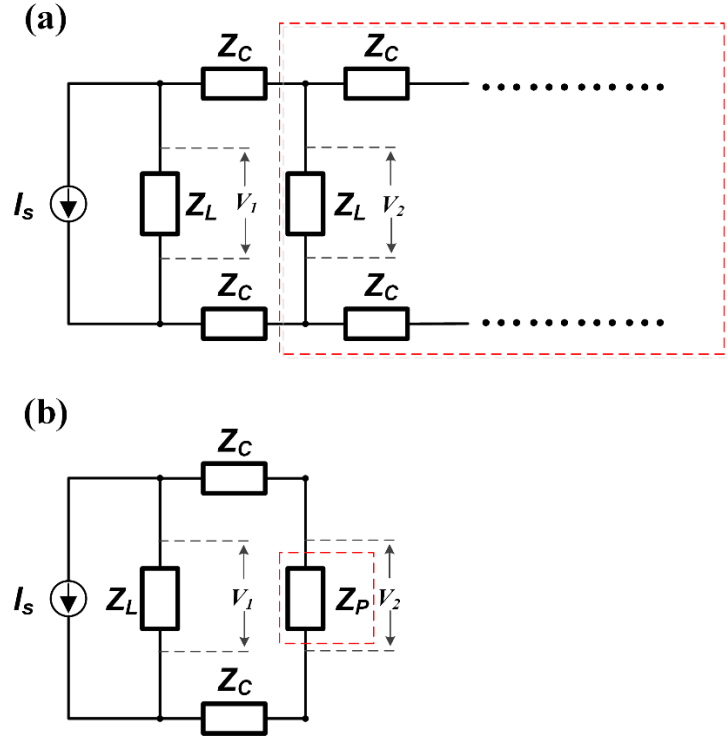

Figure1 (a) Simplified circuit cascade, $Z_L$ represents the impedance of each stage, which is a parallel RLC circuit in the neural circuit of single axon; (b) The circuit cascade in the red dash box in (a) is considered as block in the red dash box in (b) with impedance of $Z_P$.

The circuit of the single axon can be simplified as a circuit cascade as shown in Figure1 (a). The load of each stage is $Z_L$ and the impedance of the connection between each stage is $Z_C$.

A current source $I_S$ is connected with the first stage and the voltage amplitude of $n$th stage is denoted as $V_n$. Then define the decay constant $\lambda$ as:

$$\frac{V_{n+1}}{V_n} = \lambda$$

Here we will investigate how the $\lambda$ changed with frequency of the input current $I_S$.

The total impedance of the whole cascade will converge to a fixed value, denoted as $Z_P$. Since this is cascade is with infinite number of stages, an extra stage connected to this cascade will not affect the total impedance as shown in Figure1 (b), then the equation for $Z_P$ is:

$$Z_L \mathbin{/\mkern-5mu/} (2 \times Z_C + Z_P) = Z_P$$

Solve this equation:

$$Z_P = \sqrt{Z_C^2 + 2 \times Z_C \times Z_L} - Z_C$$

Then

$$\lambda = \frac{V_2}{V_1} = \frac{Z_P}{Z_P + 2 \times Z_C} = \frac{\sqrt{Z_C^2 + 2 \times Z_C \times Z_L} - Z_C}{\sqrt{Z_C^2 + 2 \times Z_C \times Z_L} + Z_C}$$

Set $\frac{Z_L}{Z_C} = \alpha$; then

$$\lambda = \frac{\sqrt{1 + 2\alpha} - 1}{\sqrt{1 + 2\alpha} + 1} = 1 - \frac{2}{\sqrt{1 + 2\alpha} + 1}$$

In this equation, $\lambda$ increases monotonically with increasing $\alpha$, and $Z_C$ is a constant value here, so $\lambda$ increases monotonically with increasing $Z_L$. In other words, a higher load, $Z_L$, results in a lower decay, which is a higher $\lambda$.

Since $Z_L$ represents the impedance of a parallel RLC circuit, which reaches the maximum value at its resonance frequency, the signal propagation will have a minimum decay when its frequency is the resonance frequency of the parallel RLC circuit as each stage.